# Privacy Preserving Machine Learning: Threats and Solutions

Mohammad Al-Rubaie | Iowa State University

J. Morris Chang | University of South Florida

**Abstract:** For privacy concerns to be addressed adequately in today's machine learning systems, the knowledge gap between the machine learning and privacy communities must be bridged. This article aims to provide an introduction to the intersection of both fields with special emphasis on the techniques used to protect the data.

**Keywords:** privacy, machine learning, differential privacy, cryptography, dimensionality reduction.

## Introduction

Our search queries, browsing history, purchase transactions, the videos we watch, and our movies' preferences are but few types of information that are being collected and stored on daily basis. This data collection happens within our mobile devices and computers, on the streets, and even in our own offices and homes. Such private data is being used for a variety of machine learning applications.

Machine learning (ML) is being increasingly utilized for a variety of applications from intrusion detection to recommending new movies. Some ML applications require private individuals' data. Such private data is uploaded to centralized locations in clear text for ML algorithms to extract patterns, and build models from them. The problem is not limited to the threats associated with having all this private data exposed to insider threat at these companies, or outsider threat if the companies holding these data sets were hacked. In addition, it is possible to glean extra information about the private data sets even if the data was anonymized[1], or the data itself and the ML models were inaccessible and only the testing results were revealed[2].

In this article, we will describe machine learning tasks and applications, and the potential threats associated with current methods of collecting data or building ML systems. We will further elaborate on the techniques proposed to protect the privacy of individuals or corporates. Our intention is to bridge the gap between machine learning and privacy and security technologies by helping professionals of either fields to be more acquainted with machine learning, the potential threats to privacy, the proposed solutions, and the challenges that lie ahead.

## Machine Learning

Arthur Samuel, a pioneer in the fields of computer gaming and artificial intelligence, described machine learning as "a field of study that gives computers the ability to learn without being explicitly programmed". The aim of machine learning algorithms is to learn how to perform certain tasks by generalizing from data. Such tasks might include giving accurate predictions or finding structures in data.

The input data to a ML algorithm is usually represented as a set of samples. Each sample would contain a set of feature values. For example, consider a 100x100 pixels photo, where each pixel is represented by a single number (0-255 grayscale). We can use these pixel values to form a vector of length 10,000, which is normally called a feature vector. Each photo, represented as a feature vector, can be associated with a label (e.g., name of the person in the photo). An ML algorithm would use a training set formed of multiple feature vectors, and their associated labels, to build an ML model. This process is called the training or learning phase. When presented with a new *test* sample, this ML model should give the predicted label (person's name or identifier in face recognition applications). The ability of such an ML model to accurately predict the label is a measure of how well this ML model generalizes to unseen data. It is measured empirically by the test error (generalization error), and it can depend on the quality and quantity of the data used for training the model, what ML algorithm was used to build the model, the selection of ML algorithm hyper parameters (e.g., using cross validation), and even the features' extraction method (if any was required).

In general, some feature extraction method might be needed to produce useful features from the raw data, such as the preprocessing step for pictures (as raw data) which might involve face detection, followed by cropping and resizing to 100x100 pixels to match the feature vector length[3], or projecting the data to lower dimensions using PCA[4]. Feature engineering utilizes domain knowledge to produce features from raw data, and it is important for many applications, however, in many modern applications it has been reduced to mere preprocessing steps. Data sets are generally formed of feature vectors, regardless of whether this data is labeled or not, which depends on the application or learning style.

In general, we can categorize ML algorithms based on their learning style into supervised or unsupervised learning (or a combination of both):

*Supervised Learning:* that utilizes labeled data where each feature vector is associated with an output value that might be a class label (classification), or a continuous value (regression). This labeled data is used to build models (training phase) that can predict the label of new feature vectors (testing phase).

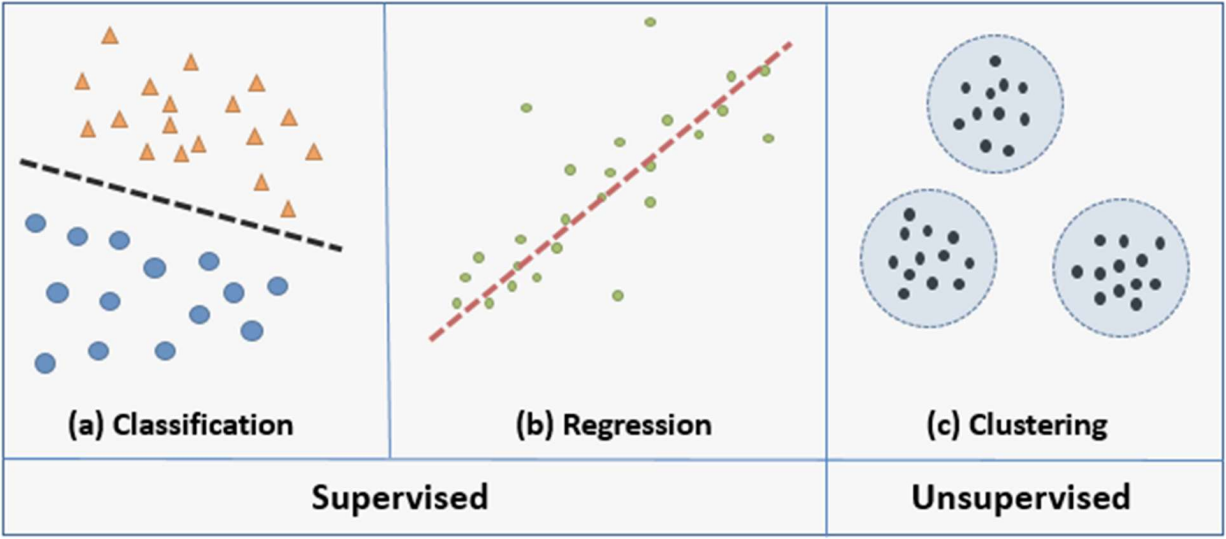

Figure 1: Machine Learning Tasks Overview. (a) Classification: finding a separating dashed line between the two classes, "circles" and "triangles" classes, (b) Regression: fitting a predictive model (dashed line to the observed data points, (c) Clustering: grouping a set of samples into a number of clusters.

With *classification*, the samples (feature vectors) belong to two or more classes, and the objective of the ML algorithm is to determine the class to which the new sample belongs. Some algorithm might achieve that by finding a separating hyperplane between the different classes as can be seen in fig. 1a. An example application is face recognition where a face image can be tested to ascertain that it belongs to a certain person. Multiple classification algorithms can be used for each of the above applications such as Support Vector Machines (SVM), Neural Networks or Logistic Regression.

When the label of a sample is a continuous value (also called the dependent or response variable) rather than a discrete one, the task is called *regression*. The samples are formed of features that are also called independent variables. The target of regression would be fitting a predictive model (such as a line) to an observed data set such that the distances between the observed data points and this line are minimized (fig. 2b). A simple example would be estimating the price of a house based on its location, area, and number of rooms.

***Unsupervised Learning:*** With this type of learning, data is not labeled as feature vectors do not come with a class label or a response variable. The target in this case would be to find structure in the data. ***Clustering*** is probably the most common unsupervised learning technique, and its aim is to group a set of samples into different clusters (fig. 2c). Samples in the same cluster are supposed to be relatively similar to each other, and different from samples in other clusters (the similarity measure could be the Euclidean distance). K-means is one of the most popular clustering methods.

***Semi-supervised Learning:*** Labeling the data can be expensive as it requires human experts or special devices; hence, only some of data gets labeled sometimes, while the vast majority remains unlabeled. Researchers found that even having a small portion of labeled data can considerably improve the learning process.

***Other Applications:*** Some ML tasks and applications do not strictly fall into one of the categories above as they can be performed either in a supervised or unsupervised way. Examples include dimensionality reduction and recommender systems.

## Threats

In each of the ML tasks mentioned above, three different roles are possible: the input party (data owners or contributors), the computation party and the results' party[5]. In such systems, the data owner(s) send their data to the computation party that performs the required ML task and delivers the output to the results' party. Such output could be an ML model that the results' party can utilize for testing new samples. In other cases, the computation party might keep the ML model, and performs the task of testing new samples submitted by the results' party, and returning the testing results to the results' party. If all three roles are assumed by the same entity, then privacy is naturally preserved; however, when these roles are distributed across two or more entities, then privacy enhancing technologies are needed.

It is common to have the same entity be both the computation and the results' parties, and this entity is mostly separate from the data owners. In fact, with all of the data that is collected from individuals around the world on daily basis, data owners might not be aware of how the data collected from them is being used (or misused), and in many cases, not even aware that some data types are being collected.

There are multiple levels of threats depending on the privacy leaks associated with the data sharing process as can be seen from Fig. 2, and the following paragraphs:

***Private Data in the Clear:*** If the data owner(s) are separate from the computation party, then the private data would be transferred to the computation party, possibly over a secure channel. However, it would most likely reside in the computation server(s) in its original form, i.e. not encrypted or transformed in any way. This is the biggest type of threat as the private data would be susceptible to both insider and outsider attacks. Such private data could be stored as raw data, or as features extracted from the raw data. Naturally, storing it in a raw form imposes greater threat as the data is ready to be processed in any way possible.

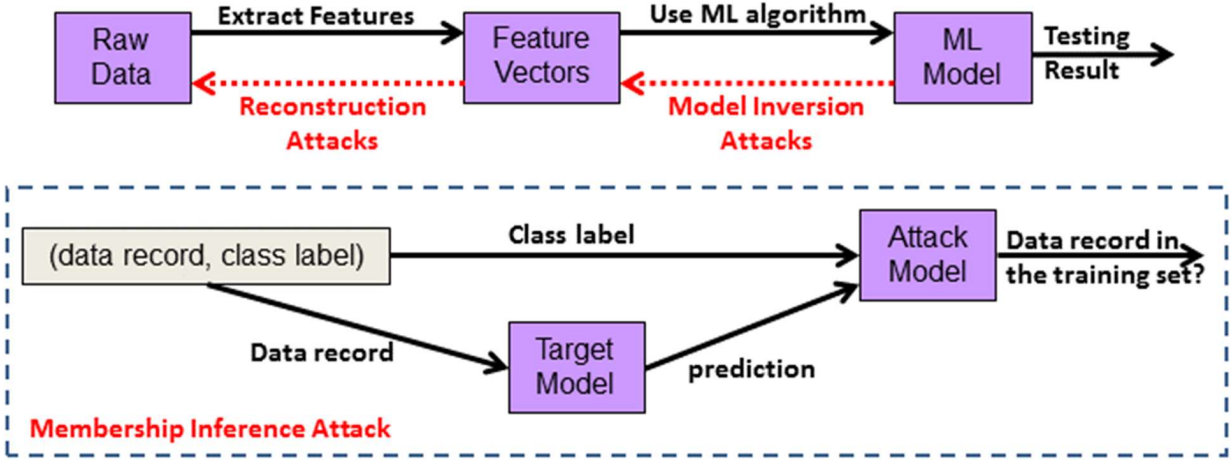

Figure 2: Machine Learning and threats

***Reconstruction Attacks:*** Even when only the features (extracted from the raw data) are transferred to, and stored in the computation party server(s), there is a threat imposed by reconstruction attacks. In this case, the ***adversary's goal*** is reconstructing the raw private data by using their ***knowledge*** of the feature vectors. Reconstruction attacks require ***white-box*** access to the ML model, i.e. the feature vectors in a model must be known. Such attacks could be possible when the feature vectors used for the ML training phase were not deleted after building the desired ML model. In fact, some ML algorithms such as SVM or kNN store feature vectors in the model itself. Examples of successful cases of reconstruction attacks include: fingerprint reconstruction[6] where a fingerprint image (raw data) could be reconstructed from a minutiae template (features), and mobile device touch gesture reconstruction[7] where touch events (raw data) were reconstructed from gesture features such as velocity and direction. In both cases, a privacy threat (caused by not protecting the private data in its feature form) resulted in a security threat to authentication systems that in turn results in failure to protect the data owners' privacy (since attackers might gain access to the data owners' devices). While the aim of these cases was misguiding a ML system into thinking the reconstructed raw data belonged to a certain data owner, other reconstruction attacks might reveal private data directly such as location or age.

To resist reconstruction attacks, ML models that store explicit feature vectors (e.g. SVM) should be avoided, or, if used, they should not be provided to the results' party. Moreover, protection against model inversion attacks should be in place to prevent synthesizing feature vectors.

***Model Inversion Attacks:*** Some ML algorithms produce models where explicit feature vectors are not stored in the ML model, e.g. ridge regression or neural networks. Hence, the ***adversary's knowledge*** would be limited to either: (a) an ML model with no stored feature vectors (white-box access), or (b) only the responses returned by the computation party when the results' party submits new testing samples (black-box access). Here, the ***adversary's target*** is creating feature vectors that resemble those used to create an ML model by utilizing the responses received from that ML model. Such attacks utilize the confidence information (e.g. probability or SVM decision value) that are sent back as a response for testing samples submitted by the results' party. These attacks produce an average that represents a certain class; hence, they would be most threatening to privacy when a certain class represents a single individual, as in face recognition. It should be noted that Fredrikson *et al.*[2] demonstration of their model inversion attack also included reconstruction in the same step. This was because the features in their case matched the raw data (face images).

To resist such attacks, the results' party should be limited to black-box access, and the output should be limited; thereby decreasing the black-box adversary's knowledge. For example, the attack success rate decreased when classification algorithms reported rounded confidence values[2] or just the predicted class labels[7]. A step further could be aggregating the result of testing multiple samples[7], but this approach would not be appropriate for all applications.

***Membership Inference Attacks:*** While model inversion attacks do not produce an actual sample from the training set, nor do they infer whether a sample was in the training set based on the ML model output, membership inference attacks do the latter. Given an ML model and a sample (***adversary's knowledge***), membership inference attacks aim to determine if the sample was a member of the training set used to build this ML model (***adversary's target***). This attack could be used by an adversary to learn whether a certain individual's record was used to train an ML model associated with a specific disease. Such attacks utilize the differences in the ML model predictions on samples that were used in the training set versus those that were not included. Shokri *et al.*[8] investigated these attacks, and trained attack models that take a sample's correct label and the target ML model prediction as inputs, and determines whether the sample was in the training set or not. These attack models were trained using shadow models built from data generated using three methods: model inversion attack, statistics-based synthesis, or noisy real data. While training the attack models utilized a black or white-box access, performing the attacks using these models only required a black-box adversary.

Shokri *et al.*[8] tried different mitigation techniques such as regularization and coarse precision of prediction vectors, and found that limiting the output to the class label was the most effective, albeit not enough to thwart the attack completely. By definition, differential privacy can resist membership inference attacks (as will be seen in the next section).

*De-anonymization (Re-Identification):* Anonymization by removing personal identifiers before releasing the data to the public may seem like a natural approach for protecting the privacy of individuals. Indeed, some companies attempted to protect their users' privacy by only releasing anonymized versions of their datasets, as was the case with the anonymous movie ratings released by Netflix to aid contestants for its 1M$ prize to build better recommender systems (for movies). Despite the anonymization, researchers were able to utilize this dataset along with IMDB background knowledge to identify the Netflix records of known users, and were further able to deduce the users' apparent political preferences[1]. This incident demonstrates that anonymization cannot reliably protect the privacy of individuals in the face of strong adversaries.

## Privacy-Preserving Machine Learning (PPML)

Many privacy-enhancing techniques concentrated on allowing multiple input parties to collaboratively train ML models without releasing their private data in its original form. This was mainly performed by utilizing cryptographic approaches, or differentially-private data release (perturbation techniques). Differential privacy is especially effective in preventing membership inference attacks. Finally, as discussed above, the success of model inversion and membership inference attacks can be decreased by limiting the model prediction output (e.g. class labels only).

### Cryptographic Approaches

When a certain ML application requires data from multiple input parties, cryptographic protocols could be utilized to perform ML training/testing on encrypted data. In many of these techniques, achieving better efficiency involved having data owners contribute their encrypted data to the computation servers, which would reduce the problem to a secure two/three party computation setting. In addition to increased efficiency, such approaches have the benefit of not requiring the input parties to remain online.

Most of these approaches address the case of horizontally-partitioned data: Each data owner has collected the same set of features for different data objects. Face recognition is one example since each person that wants an ML model trained for her face can submit multiple feature vectors extracted from their own photos. In each of these cases, the same set of features is extracted by each data owner.

Homomorphic encryption, garbled circuits, secret sharing and secure processors are the most widely used cryptographic techniques to achieve PPML:

***Homomorphic Encryption:*** Fully homomorphic encryption enables the computation on encrypted data, with operations such as addition and multiplication that can be used as basis for more complex arbitrary functions. Due to the high cost associated with frequently bootstrapping the cipher text (refreshing the cipher text because of the accumulated noise), additive homomorphic encryption schemes were mostly used in PPML approaches. Such schemes only enable addition operations on encrypted data, and multiplication by a plaintext. A popular example is Paillier cryptosystem.

To extend additive homomorphic encryption functionality, protocols were developed to enable comparison of two encrypted values, or to perform secure multiplication and decryption operations, mostly by "blinding" the cipher text through adding an encrypted random value to the encrypted value that needs to be protected. To increase the efficiency of using additive homomorphic encryption, data packing techniques were developed to enable more than one plain text value to be encrypted by the same cipher text. Such techniques were employed by some PPML approaches to enable efficient and secure PPML systems such as the collaborative filtering system[9] proposed by Erkin *et al.* which used all of the previous techniques. In this system, data owners contribute data encrypted with the public key of a privacy service provider (PSP), but send the encrypted data to the service provider (SP). The PSP provides privacy and computation services, while the SP provides storage and computation services with the intention of generating private recommendations for its customers (the data owners). For such system to be secure, the SP and the PSP must not collude. Since they provide different services, it is understandable that the SP and the PSP could be different companies; hence, the non-collusion assumption is plausible. In this system, the data owners are both input and results' parties while the SP and PSP are the computation parties.

*Garbled Circuits*: Assuming a two-party setup with Alice and Bob wanting to obtain the result of a function computed on their private inputs, Alice can convert the function into a garbled circuit, and send this circuit along with her garbled input. Bob obtains the garbled version of his input from Alice without her learning anything about Bob's private input (e.g., using oblivious transfer). Bob can now use his garbled input with the garbled circuit to obtain the result of the required function (and can optionally share it with Alice). Some PPML approaches combined additive homomorphic encryption with Garbled circuits. Nikolaenko *et al.*[10] developed a PP ridge regression system that utilized both techniques, where an evaluator, similar to the SP in Erkin *et al.*[9], adds the encrypted shares submitted by multiple data owners to obtain encrypted intermediate values. These shares are encrypted using additive homomorphic encryption with the public key of the crypto service provider CSP (similar to the PSP in Erkin *et al.*[9]). The CSP then creates a garbled circuit and sends it to the evaluator that also obtains the garbled version of the intermediate shares from the CSP. The evaluator can proceed with using the garbled circuit and its garbled input to create the ML model(s) it needs.

Some PPML approaches concentrated on the classification task alone (testing phase rather than training/testing phases). Bost *et al.*[13] developed cryptographic building blocks using homomorphic encryption and garbled circuits that enabled them to construct three popular classification protocols: hyperplane decision, Naïve Bayes, and decision trees. The aim was to enable testing new samples while protecting both the ML model and the submitted samples.

*Secret Sharing* is a method for distributing a secret among multiple parties, with each one holding a "share" of the secret. Individual shares are of no use on their own; however, when the shares are combined, the secret can be reconstructed. With threshold secret sharing, not all the "shares" are required to reconstruct the secret; but only "t" of them ("t" refers to threshold).

In one setting, multiple input parties can generate "shares" of their private data, and send these shares to a set of non-colluding computation servers. Each server could compute a "partial result" from the "shares" it received. Finally, a results' party (or a proxy) can receive these partial results, and combine them to find the final result. Since these computation servers have similar functionalities (unlike SP and PSP mentioned above), special attention should be paid to where such servers are hosted, and which entities control them, in order to convince data owners that the computation servers would not collude. In general, secret sharing approaches might be more efficient than the other cryptographic approaches, and this resulted in having multiple commercial products that utilize secret sharing. An example is ShareMind developed by Cybernetica which was used to develop a privacy-preserving system for performing PCA computation by adapting parallelized PCA computation method to the secret sharing paradigm[11].

In another settings[12], the "shares" are distributed among other input parties (users), rather than with the computation party (server). Bonawitz *et al.*[12] developed a protocol for securely computing sums of vectors, to aggregate user-provided model updates for training a neural network model. Each user applies double-masking (blinding) on its private update vector using: a user-specific secret value, and secret values shared with other users (generated using Diffie-Hellman key agreement). To account for users dropping out before finishing the protocol, each user distributes "shares" of its user-specific secret and its Diffie-Hellman private key to other users. The server is tasked with routing messages between the users, and computing the final result if at least "t" of the users survive until the last round (threshold secret sharing). This protocol is communication-efficient (no more than twice the plain text counterpart); which makes it suitable for high-dimensional vectors.

***Secure Processors:*** While initially introduced to ensure the confidentiality and integrity of sensitive code from unauthorized access by rogue software at higher privilege levels, Intel SGX-processor are being utilized in privacy-preserving computation. Ohrimenko *et al.*[14] developed a data oblivious ML algorithms for neural networks, SVM, k-means clustering, decision trees and matrix factorization that are based on SGX-processors. The main idea involves having multiple data owners collaborate to perform one of the above mentioned ML tasks with the computation party running the ML task on an SGX-enabled data center. An adversary can control all the hardware and software in the data center except for the SGX-processors used for computation. In this system, each data owner independently establishes a secure channel with the enclave (containing the code and data), authenticates themselves, verifies the integrity of the ML code in the cloud, and securely uploads its private data to the enclave. After all the data is uploaded, the ML task is run by the secure processor, and the output is sent to the results' parties over secure authenticated channels.

**Perturbation Approaches**

Differential privacy (DP) techniques resist membership inference attacks by adding random noise to the input data, to iterations in a certain algorithm, or to the algorithm output. While most DP approaches assume a trusted aggregator of the data, local differential privacy allows each input party to add the noise locally; thus, requiring no trusted server. Finally, dimensionally reduction perturbs the data by projecting it to a lower dimensional hyperplane to prevent reconstructing the original data, and/or to restrict inference of sensitive information.

***Differential Privacy (DP):*** A randomized algorithm *M* is $\varepsilon$-differentially private if for all *S* in the range of *M*, and for all data sets *D* and *D'* differing in one record:

$$\Pr[M(D) \in S] \leq \exp(\varepsilon) \Pr[M(D') \in S]$$

Hence, DP ensures that any sequence of outputs (response to queries) is *essentially* equally likely to happen, whether a certain record was included in the data set or not[15] ("essentially" is captured by the parameter $\varepsilon$). In recent practical publications[4][16], the parameter $\varepsilon$ was set to be a single digit (smaller values indicate better privacy).

Composition is an important property of DP which enables the design and analysis of complex DP algorithms from simpler DP building blocks. The composition of a sequence of *k* mechanisms, where the $i^{th}$ mechanism provides $\varepsilon_i$-DP, is ($\sum_{i=1}^{k} \varepsilon_i$)-DP (refer to Dwork and Roth[15] for proofs and more advanced composition theorems). It is not unnatural that the strength of the privacy guarantee would degrade with repeated use of the mechanism, but DP provides a way to quantify the privacy loss.

Adding $\delta$ to the right-hand-side of the above equation yields ($\varepsilon, \delta$)-differential privacy (a relaxation with weaker privacy guarantees than pure differential privacy, i.e. $\varepsilon$-DP). The ($\varepsilon, \delta$)-DP definition was proposed to capture the privacy guarantees of the Gaussian mechanism, and due to the applications of advanced composition theorems. To avoid compromising the privacy, the value of $\delta$ has to be less than the inverse of any polynomial in the size of the database[15]. Recently, alternative relaxations of $\varepsilon$-DP were proposed, e.g. Rényi Differential Privacy (RDP)[17]. While accommodating the analysis of the Gaussian mechanism and advanced composition theorems, RDP could be preferred over ($\varepsilon, \delta$)-DP since RDP does not allow a total breach of privacy with no residual uncertainty[17].

By definition, differential privacy (DP) can deter membership inference attacks. DP is also utilized by some distributed learning approaches to enable protection of the original data in case of multiple input parties. In addition, DP is immune to post-processing; meaning that even in the presence of auxiliary information, an adversary cannot increase the privacy loss. Thus, DP neutralizes linkage attacks used for de-anonymization.

From the definition above, it's clear that DP algorithms are randomized. They can be categorized according to where, and how, the randomness is applied:

1. **Input Perturbation**: In this case, noise is added to the data itself, and after the desired non-private-computation is performed on the noisy input, the output would be differentially-private. Taking PCA as an example where Eigen-decomposition is performed

   on the covariance matrix, Dwork *et al.*[18] adds symmetric Gaussian noise matrix to the covariance matrix before performing Eigen-decomposition. The output would be a DP-projection matrix (the target here is not releasing the projected data, but the DP-projection matrix).
2. **Algorithm Perturbation:** Another approach is perturbing intermediate values in iterative algorithms. For example, Eigen-decomposition for PCA can be performed using the power method, which is an iterative algorithm. Hardt and Price[19] proposed adding Gaussian noise in each iteration of the algorithm which operates on the non-perturbed covariance matrix, leading to DP-PCA. Similarly, Abadi *et al.*[4] proposed a DP-Deep learning system by modifying the stochastic gradient descent algorithm to have Gaussian noise added in each of its iterations.
3. **Output Perturbation** involves running the non-private-learning algorithm, and then adding noise to the generated model. For cases where adding noise to the computed output would destroy its value, the Exponential mechanism could be used. Given some utility function u(D,r) that tells us how good an output r is on database D, the exponential mechanism selects an output r with probability proportional to this utility function (i.e., the output would be biased towards ones with higher quality). This method was used for achieving DP-PCA by sampling a random k-dimensional subspace that approximates the top-k PCA subspace[20].
4. Finally, ***objective perturbation*** entails adding noise to the objective function for learning algorithms such as Empirical Risk Minimization[21].

The approaches mentioned above work on data hosted by a single server (a trusted server). To enable training on disjoint datasets held by multiple input parties, Papernot *et al.*[16] proposed to first learn an ensemble of teacher models from the disjoint datasets, use these teachers to make noisy predictions on public data, which can be used in turn to build a student model. The privacy loss is determined by the number of queries made to the teachers during the student training, and do not increase as end-users query the deployed student model[16]. This approach is not limited to a single ML algorithm, but it requires adequate data quantity at each location.

***Local Differential Privacy:*** When the input parties do not have enough information to train a ML model, it might be better to utilize approaches that rely on *local* differential privacy (LDP). With LDP, each input party would perturb their data, and only release this obscure view of the data. An old, and well-known version of local privacy is randomized response (Warner 1965), which provided plausible deniability for respondents to sensitive queries. For example, a respondent would flip a fair coin: (a) if "tails", the respondent answers truthfully, and (b) if "heads", then flip a second coin, and respond "Yes" if heads, and "No" if tails. This version of randomized response (RR) is (ln 3)-differentially private[15].

RAPPOR [22] is a technology for crowdsourcing statistics from end-user client software by applying RR to Bloom filters with strong $\varepsilon$-DP guarantees. RAPPOR is deployed in Google Chrome web browser, and it permits collecting statistics on client-side values and strings, such as their categories, frequencies, and histograms. By performing RR twice with a memoization step in between, privacy protection is maintained even when multiple responses are collected from the same participant over time [22].

A ML oriented work, AnonML[23], utilized the ideas of RR for generating histograms from multiple input parties. AnonML utilizes these histograms to generate synthetic data on which a ML model can be trained. Like other local DP approaches, AnonML is a good option when no input party has enough data to build a ML model on their own (and there is no trusted aggregator).

**Dimensionality Reduction (DR)** perturbs the data by projecting it to a lower dimensional hyperplane. Such transformation is lossy, and it was suggested by Liu *et al.*[24] that it would enhance the privacy, since retrieving the exact original data from a reduced dimension version would not be possible (the possible solutions are infinite as the number of equations is less than the number of unknowns). Hence, Liu *et al.*[24] proposed to use a random matrix to reduce the dimensions of the input data. Since a random matrix might decrease the utility, other approaches used both unsupervised and supervised DR techniques such as principal component analysis (PCA), discriminant component analysis (DCA), and multidimensional scaling (MDS). These approaches try to find the best projection matrix for utility purposes, while relying on the reduced dimensionality aspect to enhance the privacy.

Since an approximation of the original data can still be obtained from the reduced dimensions, some approaches, e.g. Jiang *et al.*[25], combined dimensionality reduction with DP to achieve differentially-private data publishing. While some entities might seek total hiding of their data, DR has another benefit for privacy. For datasets that have samples with two labels: a utility label and a privacy label, Kung[26] proposes a DR method to enable the data owner to project her data in a way that enables maximizing the accuracy of learning for the utility labels, while decreasing the accuracy for learning the privacy labels. Although this method does not eliminate all privacy risks of the data, it enables controlling the misuse of the data when the privacy target is known.

## Challenges and outlook

Despite the aforementioned techniques to protect the private data while performing ML training and/or testing, non-privacy-aware ML algorithms are still being widely used, and private data is still being uploaded to the cloud on daily basis. Current laws might force companies to declare that they are collecting data, and might even give the user the option to opt out of such data collection, but it seems like a zero or one decision. There should be an alternative option that utilizes some of privacy-preserving ML techniques already proposed by the research community, however, some issues might hinder achieving that.

The first issue is flexibility. Many of the aforementioned PPML techniques are tied to a certain ML algorithm. This would pose a problem especially considering the fact that new ideas and advances in ML are being proposed on regular basis. Thus, some of these PPML techniques might need to be re-purposed regularly to cope up with the new advances. In such cases, transformed data release, the distributed approach by Papernot *et al.*[16], or local differential privacy might be favorable as they provide the opportunity to apply new advances in ML without extensive customization.

Another issue is scalability, both in terms of processing and communication costs. A similar problem happens with algorithms design where higher processing powers was accompanied by greater amounts of data; hence, new techniques were required. New advances in ML concentrated on efficiency, parallelism and reducing the communication cost. However, many PPML techniques impose additional processing and communication costs that might limit the ability to utilize the huge amounts of data available today. Promising PPML techniques could be those that are already built around distributed processing, and only exchanging summary statistics or model parameters.

In addition, security assumptions for some PPML systems need to be addressed properly. An example is non-collusion assumptions where two or more of the computation parties might be assumed not to collude. Such assumptions might be easier to justify when the different parties perform different roles; thus implying they could be different companies (such as the service provider and the PSP in Erkin *et al.*[9]). the question remains: who will take the role of the privacy service provider? And how will they make their profit to sustain their business?

Policies are yet another issue. Privacy policies can help the data owners specify which data is being shared, according to what rules or privacy guarantees, who will use the data, and for what purpose? It is important to determine if the policy will be enforced at the client side, meaning that it would be transformed to a form that limits all other uses, or at least eliminates some uses with known privacy threats. Another option would be sending the policy with the data to the computation servers where the servers have to be trusted to honor the policy rules. If the computation party and the results' party are controlled by the same entity, trust would be an issue.

To summarize and present other issues, we can take the human data interaction (HDI) principles into consideration. While HDI applies to any type of data sharing, we concentrate here on the specific case of Machine Learning. Mortier *et al.* identify three core themes of HDI[27]: Legibility, Agency and Negotiability. **Legibility** means informing data owners that their data is being collected, what data is being collected, how it is being used (including what inferences might be made), and potentially providing legible explanation of how their privacy is being preserved. While some companies, e.g. Google and Apple, started using local differential privacy for data collection, it might still be a challenge to explain the technology to the public, and the implications of their choice of specific $\varepsilon$ on the users' privacy.

***Agency*** is concerned with enabling data owners to have control of their data in ML systems. Enabling data owners to opt out of data collection, or set rules and policies about how their data is being used, and even adjust some incorrect inferences that were made about them. One example would be machine unlearning[28] where the target was adjusting an ML model, that was built using wrong data about a data owner, and enable incremental unlearning without having to start the training process from scratch. Such techniques could be beneficial to help realize "the right to be forgotten" that was recently enforced by the European Union.

Since the definition of privacy (and associated threats) might evolve overtime, **Negotiability** in PPML systems is essential. Negotiability enables the data owners to re-evaluate their data sharing decisions (e.g. enabling them to withdraw from ML systems completely or partially, or to simply change the policies of using their data). As we design current systems, how can we make the policy definitions extensible to enable addressing unforeseen concerns in policies that were created to address today's concerns?

**Acknowledgment**

## References

1	Narayanan, Arvind and Shmatikov, Vitaly (2008) 'Robust de-anonymization of large sparse datasets'. *Proceedings - IEEE Symposium on Security and Privacy*, pp. 111–125.

2	Fredrikson, Matt, Ristenpart, Thomas, Tech, Cornell, Jha, Somesh and Ristenpart, Thomas (2015) 'Model Inversion Attacks that Exploit Confidence Information and Basic Countermeasures'. *Proceedings of the 22nd ACM SIGSAC Conference on Computer and Communications Security - CCS '15*, pp. 1322–1333.

3	Bishop, Christopher M. (2007) 'Pattern Recognition And Machine Learning'. *Book*, p. 738.

4	Abadi, Martin, Chu, Andy, Goodfellow, Ian, McMahan, H. Brendan, et al. (2016) 'Deep Learning with Differential Privacy'. *Proceedings of the 2016 ACM SIGSAC Conference on Computer and Communications Security - CCS'16*, (Ccs), pp. 308–318.

5	Bogdanov, Dan, Kamm, Liina, Laur, Sven, Pruulmann-Vengerfeldt, Pille, et al. (2014) 'Privacy-Preserving Statistical Data Analysis on Federated Databases'. *Lecture Notes in Computer Science (including subseries Lecture Notes in Artificial Intelligence and Lecture Notes in Bioinformatics)*, 8450 LNCS, pp. 30–55.

6	Jianjiang Feng and Jain, A K (2011) 'Fingerprint Reconstruction: From Minutiae to Phase'. *IEEE Transactions on Pattern Analysis and Machine Intelligence*, 33(2), pp. 209–223.

7	Al-Rubaie, Mohammad and Chang, J. Morris (2016) 'Reconstruction Attacks Against Mobile-Based Continuous Authentication Systems in the Cloud'. *IEEE Transactions on Information Forensics and Security*, 11(12), pp. 2648–2663.

8	Shokri, Reza, Stronati, Marco, Song, Congzheng and Shmatikov, Vitaly (2017) 'Membership Inference Attacks Against Machine Learning Models', in *2017 IEEE Symposium on Security and Privacy (SP)*, IEEE, pp. 3–18.

9	Erkin, Zekeriya, Veugen, Thijs, Toft, Tomas and Lagendijk, Reginald L. (2012) 'Generating private recommendations efficiently using homomorphic encryption and data packing'. *IEEE Transactions on Information Forensics and Security*, 7(3), pp. 1053–1066.

10	Nikolaenko, Valeria, Weinsberg, Udi, Ioannidis, Stratis, Joye, Marc, et al. (2013) 'Privacy-preserving ridge regression on hundreds of millions of records'. *Proceedings - IEEE Symposium on Security and Privacy*, pp. 334–348.

11	Bogdanov, Dan, Kamm, Liina, Laur, Sven and Sokk, Ville (n.d.) 'Implementation and Evaluation of an Algorithm for Cryptographically Private Principal Component Analysis on Genomic Data – Short Paper'.

12   Bonawitz, Keith, Ivanov, Vladimir, Kreuter, Ben, Marcedone, Antonio, et al. (2017) 'Practical Secure Aggregation for Privacy Preserving Machine Learning'. *Cryptology ePrint Archive: Report 2017/281*.

13   Bost, Raphaël, Popa, Ra, Tu, Stephen and Goldwasser, S (2014) 'Machine Learning Classification over Encrypted Data'. *Eprint.Iacr.Org*, (February), pp. 1–31.

14   Ohrimenko, Olga, Schuster, Felix, Fournet, Cedric, Mehta, Aastha, et al. (2016) 'Oblivious Multi-Party Machine Learning on Trusted Processors'. *25th USENIX Security Symposium (USENIX Security 16)*, pp. 619–636.

15   Dwork, Cynthia and Roth, Aaron (2014) 'The Algorithmic Foundations of Differential Privacy'. *Foundations and Trends in Theoretical Computer Science*, 9(2013), pp. 211–407.

16   Papernot, Nicolas, Abadi, Mart'\in, Erlingsson, Úlfar, Goodfellow, Ian and Talwar, Kunal (2016) 'Semi-supervised knowledge transfer for deep learning from private training data'. *arXiv preprint arXiv:1610.05755*.

17   Mironov, Ilya (2017) 'Rényi Differential Privacy', in *2017 IEEE 30th Computer Security Foundations Symposium (CSF)*, IEEE, pp. 263–275.

18   Dwork, Cynthia, Talwar, Kunal, Thakurta, Abhradeep and Zhang, Li (2014) 'Analyze gauss: Optimal Bounds for Privacy-preserving Principal Component Analysis'. *Stoc*, pp. 11–20.

19   Hardt, Moritz and Price, Eric (2014) 'The Noisy Power Method : A Meta Algorithm with Applications', in *Advances in Neural Information Processing Systems*, pp. 2861–2869.

20   Chaudhuri, Kamalika, Sarwate, Anand D and Sinha, Kaushik (2013) 'A near-optimal algorithm for differentially-private principal components'. *The Journal of Machine Learning Research*, 14(1), pp. 2905–2943.

21   Chaudhuri, Kamalika, Monteleoni, Claire and Sarwate, Anand D. (2011) 'Differentially Private Empirical Risk Minimization'. *The Journal of Machine Learning Research*, 12, pp. 1069–1109.

22   Erlingsson, Úlfar, Pihur, Vasyl and Korolova, Aleksandra (2014) 'RAPPOR: Randomized Aggregatable Privacy-Preserving Ordinal Response', in *Proceedings of the 2014 ACM SIGSAC Conference on Computer and Communications Security - CCS '14*, New York, New York, USA, ACM Press, pp. 1054–1067.

23   Cyphers, Bennett and Veeramachaneni, Kalyan (n.d.) 'AnonML: Locally private machine learning over a network of peers'.

24   Liu, Kun, Kargupta, Hillol and Ryan, Jessica (2006) 'Random projection-based

multiplicative data perturbation for privacy preserving distributed data mining'. *IEEE Transactions on Knowledge and Data Engineering*, 18(1), pp. 92–106.

25  Jiang, Xiaoqian, Ji, Zhanglong, Wang, Shuang, Mohammed, Noman, et al. (2013) 'Differential-private data publishing through component analysis'. *Transactions on data privacy*, 6(1), p. 19.

26  Kung, S.Y. (2017) 'Compressive Privacy: From Information\/Estimation Theory to Machine Learning [Lecture Notes]'. *IEEE Signal Processing Magazine*, 34(1), pp. 94–112.

27  Mortier, Richard, Haddadi, Hamed, Henderson, Tristan, McAuley, Derek and Crowcroft, Jon (2015) 'Human-Data Interaction: The Human Face of the Data-Driven Society'.

28  Cao, Yinzhi and Yang, Junfeng (2015) 'Towards making systems forget with machine unlearning'. *Proceedings - IEEE Symposium on Security and Privacy*, 2015–July, pp. 463–480.

**Mohammad Al-Rubaie** is pursuing the Ph.D. degree in computer engineering at Iowa State University, Ames, Iowa. He is currently a predoctoral Intern at the University of South Florida. His research interests include: cyber security, privacy-enhancing technologies and machine learning. Contact him at mti@iastate.edu.

**J. Morris Chang** is a professor at the Electrical Engineering department in the University of South Florida. His research interests include: cyber security, wireless networks, and energy efficient computer systems. Chang received his Ph.D. degree from North Carolina State University. Contact him at chang5@usf.edu.